\documentclass[sigconf]{acmart}

\usepackage{balance}
\usepackage{graphicx}
\usepackage{threeparttable}
\usepackage{framed}
\usepackage{enumitem}
\usepackage[skip=3pt]{caption} 

\usepackage{xspace}
\xspaceaddexceptions{’}

\newif\ifanon

\ifanon
  \newcommand{\provena}{CodeTrace\xspace}
\else
  \newcommand{\provena}{Provena\xspace}
\fi

\title{``I never knew what Provena did’’: Student Perceptions and Mental Models of Authorship Tracking in CS1}

\author{Keith Tran}
\email{ktran24@ncsu.edu}
\orcid{0009-0006-2591-2227}
\affiliation{%
  \institution{North Carolina State University}
  \city{Raleigh}
  \state{NC}
  \country{USA}
}

\author{Danielle Brown}
\email{dmbrow28@ncsu.edu}
\orcid{0009-0001-1668-2232}
\affiliation{%
  \institution{North Carolina State University}
  \city{Raleigh}
  \state{NC}
  \country{USA}
}

\author{Thomas W. Price}
\email{twprice@ncsu.edu}
\orcid{0000-0001-9375-2292}
\affiliation{%
  \institution{North Carolina State University}
  \city{Raleigh}
  \state{NC}
  \country{USA}
}

\copyrightyear{2027}
\acmYear{2027}
\setcopyright{cc}
\setcctype{by}
\acmConference[SIGCSE TS 2027]{Proceedings of the 58th ACM Technical Symposium on Computer Science Education V.1}{February 17--20, 2027}{Sacramento, CA, USA}
\acmBooktitle{Proceedings of the 58th ACM Technical Symposium on Computer Science Education V.1 (SIGCSE TS 2027), February 17--20, 2027, Sacramento, CA, USA}
\acmDOI{10.1145/3857222.3860138}
\acmISBN{979-8-4007-3075-7/2027/02}

\begin{document}

\begin{abstract}
As generative AI can complete most introductory programming assignments, instructors are adopting authorship-tracking tools that record how code is written, to preserve and verify students’ own work. Prior work has focused on whether these tools can detect AI-generated work, with less attention to how students being tracked understand and respond to them. We surveyed students in an introductory programming course for non-computing majors and analyzed their open-ended responses about one such tool, Provena, using codebook thematic analysis. Students held incomplete mental models of how the tool worked, even after instructors introduced it, and responded to what they believed it was doing rather than what it did. Most students did not mind being recorded, and some valued it as proof of their work, while many feared it would misread their normal coding habits as cheating. We argue that whether authorship tracking helps or harms learning may depend less on what the tool does than on the mental model students hold, so how instructors explain these tools, not just which tool they adopt, can shape how students experience them.
\end{abstract}



\maketitle

\section{Introduction}
\label{sec:intro}

Learning to program requires practice. Students learn by working through errors, writing and testing code, and trying different solutions, and this active practice, which prior work calls the doer effect, drives learning more than watching or reading \cite{koedinger2016doer}. Generative AI now puts that practice at risk, as tools such as ChatGPT can solve most introductory programming problems on demand \cite{savelka2023thrilled, finnie2023my}. Because grading usually rewards the correctness of the final code rather than how it was produced, a student can submit an AI-generated solution for full credit, weakening the incentive to practice and leaving students who do the work feeling their effort no longer counts when classmates earn the same grades with far less of it \cite{tran2026why, prather2024widening}.

In response, a growing number of instructors are turning to tools that record how code is written. These tools capture the entire editing process and keystrokes behind a submission, a class of tools we call \textbf{authorship tracking}. Such tools are already in use, both in commercial platforms\footnote{\url{https://www.codio.com/features/code-plagiarism-detection-cheating-prevention}} and research systems \cite{kim2022watcher, syndergaard2024revisiting, he2026recap}. They aim both to deter shortcuts, protecting the incentive to practice, and to establish that a submission reflects a student's own effort so that a grade means what it should. At the same time, students and instructors may push back over their surveillance, the risk of penalizing students who did nothing wrong, and their emphasis on assessment over learning \cite{balash2021examining, gillo2026writing, fischer2024seeking}.

Authorship tracking is a new response to a new problem, and instructors have little guidance on how to frame these tools and how clearly to describe how they work. These choices may matter as much as the tool itself. The tool is present during the very practice that drives learning, so if being watched changes how students engage with that work, it can carry unintended consequences for learning. And a tool’s perceived efficacy matters as much as its actual efficacy. Even an accurate tool cannot deter shortcuts or assure students that grades are fair if they do not see it as working or as fair. 

We do not yet know how students make sense of being tracked. Research on authorship and process tracking has focused on detection~ \cite{ihantola2015educational, kim2022watcher, he2026recap}, and few studies have explored students' perceptions or whether these tools support or undermine the learning environment. 

In this study, we explore how students perceive authorship tracking, using \provena~\cite{price2026beyond} as one implementation. \provena records how a student’s code develops as they write it and flags submissions that show little of this process for an instructor to review. We surveyed students in an introductory programming (CS1) course for non-CS engineering majors. These students make up a large share of introductory programming enrollment and often take CS1 to satisfy a requirement rather than to pursue computing, so the incentive to practice, and the experience of being tracked, may be different for them than for intended majors \cite{tran2026why}. The survey used Likert and open-ended items, and we analyzed the open-ended responses with codebook thematic analysis (TA). We address the following research questions: \\ \\
\textbf{RQ1}: How do students make sense of what authorship tracking does and why instructors used it? \\
\textbf{RQ2}: How do students perceive the goals of authorship tracking, how aligned do they see themselves with those goals, and how valid and fair do they find them? \\
\textbf{RQ3}: What harms or concerns do students perceive from authorship tracking? \\

\section{Related Work}

\textit{Detecting Cheating and Authorship.} The arrival of GenAI has intensified a long-standing effort to detect unauthorized help, and the tools built for it increasingly record how students produce their work. Within computing, detection approaches vary most in the \textit{granularity} of the data they use, which Ihantola et al. place on a spectrum from coarse submissions to fine-grained keystrokes~\cite{ihantola2015educational}. While similarity tools have long compared a student’s final submission against other code to catch copying~\cite{schleimer2003winnowing, prechelt2002finding}, GenAI removes the need to copy at all. Recent work has therefore turned to other signals. Some of these signals analyze the final product, such as machine learning classifiers that distinguish human- from AI-written code~\cite{oedingen2024chatgpt} and style anomalies~\cite{denzler2024style, vahid2024towards}. Others analyze the process itself, such as timing and event frequency~\cite{kim2022watcher}, the linearity of how code is written~\cite{hellas2017plagiarism, mcdonald2025pastetrace}, paste-versus-typed text~\cite{meier2024plagiarism}, and keystroke dynamics~\cite{byun2020detecting, kundu2024keystroke}. Most recently, RECAP pairs Copilot chat logs against a fine-grained edit history so instructors can flag students with unusually high AI reliance~\cite{he2026recap}. \provena sits at the fine-grained end of this spectrum and combines several of these signals~\cite{price2026beyond}. Where most systems reduce them to a flag or classification, it keeps the full edit history of a submission and marks where each section came from, so an instructor can review the evidence directly. This work tells us whether tracking can catch cheating, but not how students experience being tracked.

\textit{Student Perceptions of Academic Monitoring.} Where students have been studied under monitoring, they often report discomfort and mistrust. Surveying students who had taken proctored online exams, Balash et al.~\cite{balash2021examining} found that most saw the monitoring of mouse movement, gaze, and browsing as unnecessary and distracting. Students report similar anxieties about text-matching tools like Turnitin, where the fear of being wrongly flagged can erode students’ trust in their instructors~\cite{gillo2026writing} and informing students about similarity detection has itself been tried as a deterrent~\cite{karnalim2022educating}. The closest prior work is a line of studies from a CS1 course that required ShowYourWork, a keystroke-logging IDE plugin. Surveying its first deployment, Hart et al. found that most students viewed the logging positively, with some forgetting it was there at all, but a small share found it invasive, and its primary negative effect was anxiety~\cite{hart2023plagiarism}. Syndergaard and Edwards then gave students more control over their data, adding options to delete portions of their keystroke history and to check their own file for evidence of plagiarism~\cite{syndergaard2024revisiting}. Reported comfort changed little. Few students used the delete option, and the plagiarism check flagged nearly every file because assignments began with pasted starter code, marking honest work as suspicious. Together, these studies show that students see keystroke logging as a means of catching cheating and report both benefits and costs, but to our knowledge no prior work has examined what students believe such a tool actually does, before being told, and how those beliefs, rather than the tool itself, shape its perceived fairness and the experience of being recorded, questions that matter all the more now that GenAI, not copied code, is the concern. 


\section{Methods}
\label{sec:methods}

\subsection{Course context and the \provena tool}
\label{sec:context}

\textit{Population and Course Context}: This study took place in an introductory Python programming (CS1) course with 158 enrolled students at a large R1 public university in the United States in Spring 2026. The course primarily attracts non-CS Engineering majors and is required for several engineering majors. It covers Python fundamentals through two lecture sections, taught by different instructors, and weekly hands-on labs. Students completed weekly lab assignments (potentially in pairs), individual homeworks, three projects, and two exams. Because our survey asked about unauthorized help, we chose not to collect demographic data, which could be identifying and discourage candid responses. Surveys in prior years suggest the course is majority male, White, and 2nd-year.

\textit{Course AI policies}: The course did not allow GenAI on programming assignments, and instructors explained the reasons behind this policy (e.g., the importance of practice). During this semester, instructors rarely filed reports of academic integrity violations (AIVs) against students who used GenAI; instead, they created rubric items that penalized code constructs not covered in class\footnote{This was a change from prior semesters, where instructors filed many AIVs in response to AI use, taking up disproportional time.}.

\begin{figure}[t]
    \centering 
    \includegraphics[width=0.8\linewidth]{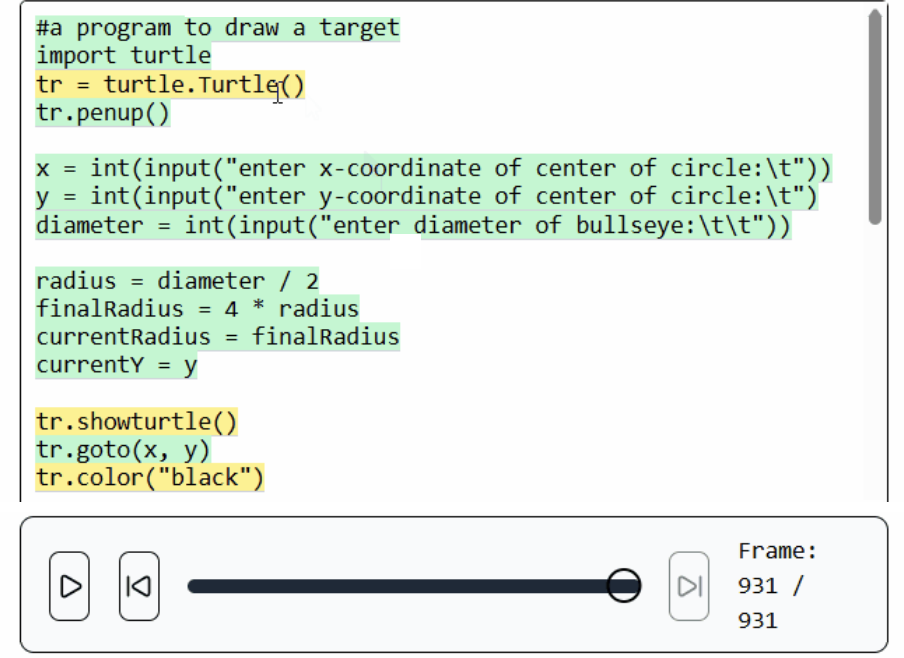}
    \Description{Screenshot of Provena's instructor view of a short Python program. Most lines are highlighted green, meaning typed, and some are highlighted yellow, meaning pasted. A playback slider below the code is at frame 931 of 931.}
    \caption{\provena's instructor view: highlighting marks each section's origin (green: typed; yellow: pasted); a slider replays how the file developed.}
    \label{fig:tool}
    \vspace{-2mm}
\end{figure}

\textit{\provena tool and deployment}: \provena is a Visual Studio Code (VS Code) extension that records keystroke, file and UI events as students write code. As a student types, it captures the edit history of each file, so an instructor can later replay how a submission was produced (Figure~\ref{fig:tool}). \provena can automatically highlight a student's submitted code with the origin of each section (e.g. manually typed, inserted by the IDE, pasted, edited outside of VS Code), and present statistics (e.g., characters per second, linearity), aiding in the detection of cheating. Instructors explained that the tool was used to ensure students did their own work and to prevent cheating. One instructor also noted that \provena tracked keystrokes and how students developed their code over time, to find unnatural patterns, such as typing code out from top to bottom with little editing.

Students were required to install and use \provena throughout the course (independent of research). Only one student reported not using it in our survey. To enforce this requirement, 10\% of each homework grade was based on whether \provena successfully collected logs for the submitted files, and some students referred to it as a \textit{keystroke requirement}. This requirement was kept minimal (only 10 event logs required) and was intended to catch logging errors (e.g., editing outside VS Code or unsynced logs), not to prevent cheating. Falling below this threshold flagged a submission for follow-up, which we refer to as the \textit{low-event flag}. The logs could later be used to inform AIVs. However, because \provena’s accuracy at detecting cheating had not yet been evaluated, in this pilot, instructors did not use it to detect AIVs or confront students. This was explicitly mentioned by one instructor but not the other. 
\subsection{Procedure}

We administered the survey through Qualtrics over two weeks near the end of the semester as an anonymous link (responses cannot be linked to individuals or lecture sections, and duplicates cannot be ruled out, though there was no incentive to repeat). To incentivize participation, students received 0.5\% extra credit (via an anonymous code). Of the 91 responses received, we excluded 18 that failed a self-reported validation item or were incomplete. That left 73 for analysis (46\% of enrolled students). This study was approved by our institution's IRB. Survey participation was voluntary, anonymous, and independent of the course requirement.

\subsection{Survey Instrument}
While some prior work has explored students' attitudes towards fine-grained authorship tracking tools like \provena \cite{hart2023plagiarism, syndergaard2024revisiting, kim2022watcher}, to our knowledge, there are no validated instruments to measure these perceptions. Our survey builds on that work, and on surveys on students' perceptions of classroom justice \cite{berti2010classroom} and employee perceptions of employer surveillance \cite{stanton2000traditional}. We add emphasis on students' understanding of the tool, its fairness, and its potential harms, with open-ended explanations of these beliefs.

The survey contained closed-ended items in five sections: 
1) students’ perceptions of \provena's purpose and their trust in the instructor (e.g., ``I understand what data \provena collects.'') (PR);
2) perceived prevalence of unauthorized help and \provena's efficacy at deterring it (UH);
3a) how well \provena supports various fairness goals in the class and
3b) overall preference to use it (VAL);
4) concerns about \provena and how it changed the experience of programming (C); and 
5) its usability (not analyzed here) (U). Each section contained three to six Likert-scale questions on a five-point scale. 

The survey also included nine open-ended questions, appearing after each set of Likert questions, asking students to explain their ratings (e.g., ``What worries, if any, did you have about your code being unfairly labeled as suspicious?’’), and achievement-goal items we did not analyze. 

Students answered the PR block, including the open-ended item asking why their instructor uses \provena and what it does, before being shown an explanation of \provena describing what it records and why. All later blocks (UH, VAL, C, U) followed this explanation, so PR responses reflect what students believed before any formal explanation from the researchers.

\subsection{Analysis Approach}
We analyzed the Likert-scale responses using descriptive statistics only, to characterize students’ perceptions and contextualize the qualitative findings. In Figure~\ref{fig:survey}, we aggregate the six fairness-goal items (3a), which showed high reliability (Cronbach's $\alpha$ = 0.89). All other analyzed items are shown individually.

\subsubsection{Qualitative Analysis}
To analyze the open-ended responses, we adopted codebook thematic analysis (TA), situated in an interpretivist (‘Big-ish Q’) tradition \cite{braun2019reflecting, braun2021one}. Our aim was to understand how students interpret and make sense of \provena, not to measure how often predefined categories appeared. Since the Likert items already capture prevalence, we do not count code frequency, and the approximate quantifiers in our Findings (``most,'' ``a few'') characterize patterns across responses rather than report counts. 

We followed the phases recommended for codebook TA \cite{braun2021one}. In familiarization, the first and second authors read all responses and wrote reflective notes in relation to our research questions. They then coded all nine open-ended items as a single data source, independently and in parallel, generating codes inductively at the meaning-unit level \cite{boyatzis1998transforming}. 

We then assembled the codebook through iterative meetings, comparing our independent codings segment by segment. Codes were merged, split, and renamed as our understanding developed, and we recorded these decisions in analytic memos. As the codes stabilized, we constructed candidate themes and refined them with feedback from the third author, building themes as patterns of shared meaning across the dataset rather than topic summaries \cite{padiyath2026reflecting}.

Consistent with this orientation, which treats researcher subjectivity as an analytic resource \cite{braun2019reflecting, braun2023doing, padiyath2026reflecting}, we do not report inter-rater agreement. Most segments converged, while a few productive disagreements were retained where informative. Quotes are reproduced verbatim, with minor spelling and grammar fixes, and square-bracketed for clarity where needed.

Because this orientation treats researchers’ perspectives as part of the analysis, we briefly describe our positionality. The first author is a Ph.D. student with qualitative experience studying how students program with AI. The second author is a CS undergraduate at the same university. The third author designed \provena to preserve incentives to practice in the age of AI (not to punish AI use), collected the data in another instructor's classroom, and served in an advisory role, shaping framing and analytic direction without coding responses. He approached the analysis focused on the potential harms of authorship tracking and student buy-in.

\section{Findings}

\begin{figure}[t]
    \centering 
    \includegraphics[width=\linewidth]{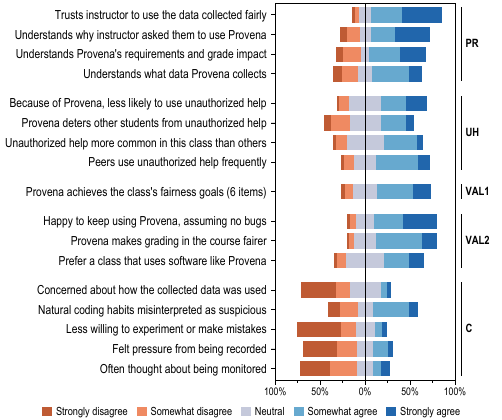}
    \Description{Diverging stacked bar chart of responses to 17 survey items in five groups, from strongly disagree to strongly agree. Agreement is highest for trusting the instructor (78\%) and being happy to keep using Provena (70\%). Agreement is lowest for the concern items (11\% to 22\%), except that 50\% agreed their natural coding habits might be misinterpreted as suspicious.}
    \caption{Distributions of responses to the analyzed survey items (n = 72--73 per item). Throughout the paper, \% agreed means somewhat or strongly agreed. Colors map each item’s five ordered options from most negative to most positive.
    }
    \label{fig:survey}
\end{figure}

Figure~\ref{fig:survey} shows the distributions of responses to the closed-ended items. Overall, students rated \provena favorably, and most would be happy to keep using it, assuming no bugs (70\%). We constructed three themes from the open-ended responses, opening each with the corresponding pattern in the Likert results. Theme 1 draws on responses to almost every open-ended item, Theme 2 mainly on the unauthorized-help and fairness items, and Theme 3 mainly on the concern items. No theme corresponds to a single research question, and we return to them in the Discussion.

\subsection{Theme 1: ``I never knew what \provena did’’: students’ partial models of \provena}

This theme captures the working mental models~\cite{norman2014some} students had of \provena's goals, how it worked, and how students cited those mental models when describing their attitudes toward and beliefs about it. Most students agreed they understood why their instructor asked them to use \provena (66\% somewhat or strongly agreed), but fewer agreed they understood what data it collects (56\%), and their open-ended responses present many different perceptions of how it worked. In most cases, these were mental models students formed on their own and largely left unquestioned.

Students were most consistent when describing their perceptions of \provena's goal. They broadly understood \provena as a way to detect cheating, ``a way for my professor to ensure my work is my own,’’ and most supported it as ``a good tool to deter cheating.’’ This support ran alongside a broader trust in the instructor, as most agreed they trusted their instructor to use the data \provena collected fairly (78\%). However, a smaller group sharply questioned the goal itself, sometimes mistrusting the school's academic integrity policies. Some questioned whether the tool was warranted at all: ``at the end of the day we are adults and we should all know not to cheat,’’ and another mistrusted the system \provena was part of, which gave ``another way to punish students who aren't doing anything wrong and have no way to prove themselves.’’

Students varied much more in \textit{how} they thought \provena accomplished its goals. Before the survey explained what \provena actually recorded, students described how they thought it worked in a number of different ways, suggesting a variety of mental models. For some, its purpose was to detect copy-and-pasted code (``I always assumed it checked if you copy and pasted or not’’), while others thought it counted keystrokes (``I think that the teacher uses \provena to count the keystrokes''). A few came closer to what the tool actually recorded, describing that it ``tracks my type history in VS code.’’ Others only learned what \provena did from the survey itself, saying plainly, ``I never knew what \provena did.’’ 

Some students believed \provena was looking for the ordinary, messy process of doing the work. One participant correctly explained, ``I knew that making mistakes and trying different things are what \provena is looking for when recording our process,” though another student feared that a cleaner and quicker solution would look suspicious: ``if I get the code right the first try, then \provena might flag it.’’ Most of these descriptions reflect some truth about \provena (it does detect pastes, keystrokes, and tries to detect AI use) but are incomplete (it does not \textit{only} use these metrics), and some are inaccurate (it does not mark correct code as suspicious). Students cited these assumptions, right, wrong, or incomplete, when describing both whether they thought \provena would work and how it felt to be recorded by it.

For some students, their assumptions about how \provena worked suggested it could easily be circumvented. The workaround they described most often was to use AI to generate the answer, then ``type it out’’ themselves rather than pasting it. One participant put it plainly, ``you can still cheat you just have to do it manually instead of copy paste.’’ A few described clearing a low-event flag directly by ``typing extra keystrokes'' or hitting ``the space bar or add/delete some characters to get the minimum keystrokes.’’ The words students used, ``manually’’, ``type it out’’, ``extra keystrokes’’, all describe an incomplete mental model in which the only thing \provena analyzes is whether keys were being pressed. In reality, manually typing an AI solution can still leave detectable traces~\cite{hellas2017plagiarism, mcdonald2025pastetrace, meier2024plagiarism}. Prior work analyzing students' logs from one project in this course found that most students whose submissions instructors later labeled as highly suspicious had typed their code rather than pasted it, but their low active time and linear editing still distinguished them~\cite{price2026beyond}.
Students also tied this model to how being recorded felt. One of the students who described clearing a flag by hitting the spacebar felt little pressure from the recording, for the same reason: ``I didn't realize \provena recorded, I thought it just counted keystrokes, so I didn't feel this pressure.'' Another with no working model likewise ``didn't really feel that pressure since I was unaware.’’ For these students, the pressure they felt, or did not feel, tracked not the tool but what they believed it was watching for. 

\textbf{Discussion of Theme 1.} Understanding what a monitoring tool is for and supporting it are not the same thing. We likely saw more buy-in for \provena’s goals in part because students trusted their instructor, and where trust is absent, explaining the tool is unlikely to produce buy-in on its own. It is worth grappling also with the fact that some students were cynical about the value of academic integrity, and sometimes with good cause. Busywork, inauthentic assignments, and a lack of buy-in for why assignments matter are among the underlying causes of cheating and legitimate student concerns~\cite{murdock2006motivational, sheard2017strategies}. An instructor adopting authorship tracking should expect some students to withhold support even when they understand its aims, and should treat trust and worthwhile assignments, not an explanation of the tool, as the groundwork for buy-in.

Our results about students’ mental models also suggest an interesting tradeoff. By keeping the discussion mostly vague (only one instructor described the mechanism), we left some students skeptical it worked at all, which may have eroded their trust in it and encouraged some to try to cheat. On the other hand, being too explicit about how the tool works and what it can catch may show students exactly how to plan around it. At the same time, a vague mental model may also lead students to worry that their legitimate coding habits could be seen as suspicious (Theme 3).

\subsection{Theme 2: ``I think \provena levels the playing field’’: whether doing your own work still counts against AI}
Likert-style questions about students' perceptions of \provena's effectiveness reveal a seeming contradiction. About half of students agreed that \provena deters their own unauthorized help (51\%), though only 37\% agreed it deters other students’, and most agreed that \provena made grading in the course fairer (67\%). At the same time, most still believed their peers used unauthorized help at least somewhat frequently (60\%), and 44\% believed it was more common in this class than in others. Our qualitative responses tell a related story. At first they aligned with the Likert data, describing \provena as fair and as keeping the class honest, but the same students who welcomed \provena as fair to those who do their own work often went on, in the next sentence, to doubt that it would change the classmates most determined to cheat. These two reactions answered the same question, whether a tool like \provena could make doing your own work still feel like it mattered when AI could otherwise complete the assignment. 

Many students valued \provena because it made doing their own work feel like it counted since ``every student has to do their own work, meaning grades are actually dependent on your knowledge and the amount that you work in the class.’’ For others, the fairness was specifically against AI, and it became most pointed when students talked about their classmates. A small number cared strongly that their peers might be cheating. One wrote that ``I think \provena levels the playing field,’’ being ``sick and tired of getting the same grade as someone who spent 10 minutes on their homework with the help of AI.’’ This student was not against AI in general: ``It's when AI does all the work for you that it becomes unfair. \provena prevents that.’’

At the same time, many students doubted \provena could deliver that fairness, because they did not think it would stop the students most determined to cheat. Some doubted the tool itself, reasoning from the incomplete model in Theme 1 that it could be worked around by retyping AI-written code by hand. Others doubted their peers rather than the tool: ``if students want to cheat, they will find ways to do it, regardless of the measures in place.’’ A few went further, treating the whole effort as pointless. One put it bluntly: ``Students will always find a way around every system so there is no point in really trying to block it. The students who want to learn will learn.’’ This was the unusual extreme. Most who doubted \provena doubted it would stop the most determined students, not that it affected themselves, which fits the gap in the ratings (51\% versus 37\%).

A few students drew a narrower conclusion, that \provena raised the cost of cheating even if it caught no one: retyping AI output is ``tedious,’’ and ``my motivation to do so drops, because [it's] now more effort [I] expend to do that than to just learn.’’ As noted in Theme 1, many of these workarounds may not be effective in practice~\cite{price2026beyond}, but without details of \provena's capabilities, and without seeing peers get caught (\provena was not used for enforcement this semester; see Section~\ref{sec:context}), students reasonably concluded it could be circumvented.

These two beliefs were less a contradiction than two ways the same goal could turn out. One student called \provena ``a good way to show that I have typed all my work out and gone back for debugging and changing things,’’ then added that ``having it or not having it wouldn't really matter since students can just work around the typing part of it.’’ 

\textbf{Discussion of Theme 2.} For many students, the value of \provena had little to do with catching anyone and much to do with restoring the sense that doing your own work still counted. When AI lets some students finish the same work in far less time, those who do their own work can feel their effort no longer matters, a resentment students voiced both here and in prior work \cite{tran2026why, hart2023plagiarism}. \provena seemed able to rebuild some of that even though it was not used for enforcement this semester, by signaling that the field was level and that the instructor valued honest work. Using it at all signaled that the instructor cared and was doing something, even to some who doubted it would stop a determined cheater. One student captured this: \provena ``mainly serves to show that the instructors make a point of us doing our own work. I respond to that in good faith and do my own work.’’ 

Some students held a black-and-white view that cheaters will cheat regardless. Others thought a sufficient deterrent could still push some toward doing the work themselves, and work on the theory of planned behavior suggests they are right \cite{skripchuk2024investigation}. Students respond to opportunities and incentives, so a tool can be effective simply by raising the cost of cheating, even when it catches no one. 
\subsection{Theme 3: ``Even though it is my own work, it just puts a slight pressure’’: students’ experience of being recorded}

Where Theme 2 focused on how students thought \provena impacted \textit{others} and the class as a whole, this theme focuses on what being recorded was actually like for students. For most, it was not something they thought much about. Few students said they often thought about being monitored (18\%) or felt pressure from being recorded (22\%), and fewer said it made them less willing to experiment or make mistakes (14\%). What students were more worried about was not being watched but being misjudged. Half were concerned that \provena might misinterpret their natural coding habits as suspicious (50\%). 

For most students, being recorded did not change much, and their indifference came in two forms. For many, it simply did not impact them because they were doing their own work and \provena had nothing to catch: it ``didn't change the way I coded at all because I wasn't using AI,’’ and ``never flagged me so I was never really worried about it because I knew I was doing the assignments fairly.’’ One student invoked the ``nothing to hide'' argument, that those unhappy with it are ``most likely unhappy with it [is] due to having something they would like to hide.’’ For others, the indifference had nothing to do with their own work. They simply did not care either way. One said plainly, ``[I] didn't know [I] was being recorded and even knowing now [I] still don’t care.’’

A smaller group experienced the recording as a benefit. They saw it as evidence of their honest work, ``if I ever got accused of cheating I could prove I didn't.’’ A few tied it to their own learning. One found that ``using AI does myself a disservice’’ and that \provena ``helps with this because it takes away the ability to use AI.’’ 

The worry students did voice was being misjudged, the highest of any concern item. Some worried \provena ``would flag it as an issue'' when they reuse their own code across assignments, others were ``scared I would use information from [a past] class that was not covered here and would get an AIV,’’ and still others worried ``my code might be labeled suspicious because it's similar to the lecture slides examples or not unique enough.’’ And for students who could not tell what would look suspicious, the uncertainty itself produced stress: ``I have no idea what looks 'suspicious' to \provena or our instructors, so naturally I believe there's more stress there.’’

For a smaller group, the discomfort was less about any particular behavior than about being watched at all. One student described how being recorded ``put the thought in the back of your head that your work can be reviewed at any time, and even though it is my own work it just puts a slight pressure for some reason.’’ Another felt it ``made me very anxious that my code would be wrongly flag[ged] and made me hyper aware and uncomfortable.’’ One said simply that they ``would try to make my programming less messy.’’ Another described a guess and check approach and was ``scared of \provena flagging it as AI’’ when ``randomly merging commands together’’ to see what would work. A third student ``felt like I needed to be right the first time.’’

\textbf{Discussion of Theme 3.} Our results suggest that the surveillance \provena requires is, in itself, acceptable to most students, which matters for the viability of authorship tracking tools. Still, a minority were uneasy, and that unease is worth addressing. Instructors can explain why the recording is necessary, and can reduce the discomfort by being clear about how the data will be used, for instance, that a recording will only be reviewed when there is specific cause for suspicion~\cite{fischer2024seeking}. They can also present the tool as something students can use to their own benefit, a way to prove their work is their own, aligning with prior work in which students wanted to show their work is genuinely theirs \cite{hart2023plagiarism, syndergaard2024revisiting}.

Novices learn by trying, failing, and revising, a process constructionism places at the center of learning to program \cite{papert2020mindstorms}. A tool that makes ``trying'' feel risky, even for a minority of students, works against the learning that the course is trying to produce. And because even a capable tool can misread honest work, and because students are often genuinely unsure what counts as a violation~\cite{liut2024didnt}, an instructor should treat a flag as the start of a conversation, not as proof. As in Theme 1, the correction may lie in the model students hold. Instructors can be upfront about what \provena records, and in particular, that the messy, exploratory process of coding is exactly what it is meant to capture. 

\vspace{-1mm}
\section{Discussion}

Our findings suggest that what a tool like \provena does to a class may depend less on what it actually does than on what students believe it does. Students made sense of \provena through mental models they built themselves, and often cited those models when describing their reactions to it (RQ1). They endorsed its goals and found it broadly fair, and described its value mostly in terms of what its presence signaled (RQ2). And its costs, like the pressure students felt, tracked what students believed it was watching for, with the most common worry being misjudged rather than being watched (RQ3). Together, these suggest instructor framing is a lever worth testing in future deployments.

\textbf{Implications for instructors}:
Prior surveys of keystroke logging found the same pattern we did, broad acceptance with an anxious minority \cite{hart2023plagiarism, syndergaard2024revisiting}. Our findings suggest one possible explanation, since students explained both their acceptance and their anxiety by what they believed the tool was doing. Authorship tracking proved viable in this course, and its presence mattered to students even though it was never used to make an accusation. But presence alone was not enough. Students still believed unauthorized AI use was common (60\%), and prior work in this course suggests they had reason to \cite{price2026beyond}, so some visible enforcement may be needed before students believe cheating is actually being stopped. Because students form their own assumptions about what they are not told (Theme 1), our themes together suggest where an instructor could draw the line. They can be explicit about what the tool records, and that the messy process of ordinary work is exactly what it is meant to record, while withholding the detection heuristics that would show a determined student how to evade it \cite{price2026beyond}.
 
\textbf{Limitations}: \provena was piloted and students experienced some sync failures, which we treat as context rather than findings. Our findings reflect one implementation of one tool in a single CS1 course at a single institution, and implementation specifics such as the low-event flag shaped what students experienced. Adopting an interpretivist framework, we aimed for transferability rather than generalizability \cite{smith2018generalizability}. Fewer than half of enrolled students responded, and they may have been more favorable toward \provena than those who did not. Despite the anonymous nature of the survey, very few students talked openly about their own use of unauthorized help, and some likely withheld their experiences of trying to outsmart the tool. \provena was not used for AIVs this semester, so we cannot cleanly separate students' doubts about the tool itself from their reaction to its not being enforced, though this unenforced first semester is also the one every adopting instructor faces first. The minimum-event requirement may also have reinforced the idea that \provena \textit{only} counted keystrokes. The two instructors also explained \provena in different levels of detail (Section~\ref{sec:context}), and because we did not record students' lecture section, we cannot compare students who heard each explanation. With self-reports collected at a single point in time, we also cannot establish that students' mental models caused their attitudes, or separate those models from their trust in the instructor.

\begin{acks}
This material is based upon work supported by the National Science Foundation under Grant \#2213792 and DGE-2137100.
\end{acks}

\newpage
\bibliographystyle{ACM-Reference-Format}
\balance
\bibliography{main}

\end{document}